# Strongly enhanced tunneling at total charge neutrality in double bilayer graphene-WSe$_2$ heterostructures

G. William Burg,[1] Nitin Prasad,[1] Kyounghwan Kim,[1] Takashi Taniguchi,[2] Kenji Watanabe,[2] Allan H. MacDonald,[3] Leonard F. Register,[1] and Emanuel Tutuc[1, *]

[1]*Microelectronics Research Center, Department of Electrical and Computer Engineering, The University of Texas at Austin, Austin, TX 78758, USA*
[2]*National Institute for Materials Science, 1-1 Namiki Tsukuba, Ibaraki 305-0044, Japan*
[3]*Department of Physics, The University of Texas at Austin, Austin, TX 78712, USA*
(Dated: April 13, 2018)

We report the experimental observation of strongly enhanced tunneling between graphene bilayers through a WSe$_2$ barrier when the graphene bilayers are populated with carriers of opposite polarity and equal density. The enhanced tunneling increases sharply in strength with decreasing temperature, and the tunneling current exhibits a vertical onset as a function of interlayer voltage at a temperature of 1.5 K. The strongly enhanced tunneling at overall neutrality departs markedly from single-particle model calculations that otherwise match the measured tunneling current-voltage characteristics well, and suggests the emergence of a many-body state with condensed interbilayer excitons when electrons and holes of equal densities populate the two layers.

In closely spaced double layer systems, interlayer electron-electron interactions can stabilize ground states that do not have a single layer counterpart. Examples include even denominator fractional quantum Hall states (QHS) in high magnetic fields at total filling factors $\nu = 1/2$ [1, 2] and $\nu = 1/4$ [3, 4], and QHSs at integer total filling factors in GaAs electron [5–7] and hole [8] double layers, and recently in graphene double layers [9, 10], that appear to host spatially indirect excitation condensates [11]. The presence of an exciton condensate is inferred from experimental signatures such as enhanced interlayer conductance [5], quantized Hall drag and dissipationless counterflow transport [6–10], and Andreev reflection [12]. Although evidence has been elusive thus far, a zero-magnetic-field counterpart of the $\nu = 1$ QHS has also been theoretically proposed [13]. Here, we investigate interlayer tunneling in double bilayer graphene heterostructures, a system which has been theoretically predicted to support a stable exciton superfluid at total charge neutrality [14, 15].

Our heterostructures consist of two rotationally aligned bilayer graphene sheets separated by bilayer WSe$_2$ with a 1.4 nm thickness [Fig. 1(a)]. The bilayer graphene crystal axes are aligned in order to enable resonant, energy and momentum conserving tunneling of carriers at the corners ($K$-points) of the graphene hexagonal Brillouin zone when the band structures of the two bilayers are aligned [16–18], characterized by a peak in the tunneling current and negative differential resistance (NDR). The resonant tunneling physics and its dependence on gate bias allows us to reliably determine the relative alignment of the bilayer band structures for samples in which the interlayer conductance is sufficiently high to prevent an independent characterization of individual layers. The heterostructures are fabricated using a layer-by-layer, dry transfer technique with the two graphene bilayers originating from a single domain to ensure matching crystal orientations [19]. The bottom and top hBN dielectrics provide atomically flat substrates, resulting in high carrier mobility [20–22]. The dielectric constants for hBN and WSe$_2$ along and perpendicular to the c-axis are $\varepsilon_{hBN}^{\parallel} = 3.0$, $\varepsilon_{hBN}^{\perp} = 6.9$ [23], $\varepsilon_{WSe_2}^{\parallel} = 7.2$ [24], and $\varepsilon_{WSe_2}^{\perp} = 14$ [25].

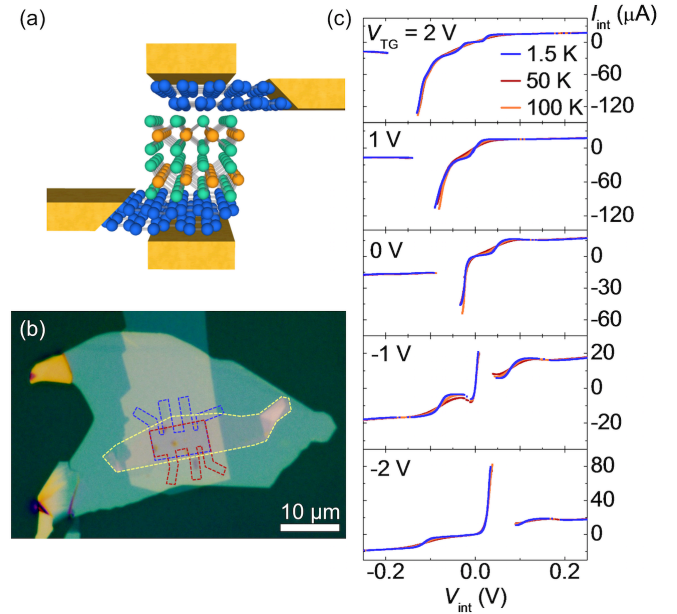

FIG. 1. (a) Schematic of a double bilayer graphene bilayer WSe$_2$ heterostructure, with top and back gates and independent contacts to each graphene bilayer. (b) Optical micrograph of a heterostructure encapsulated in hBN. Dashed lines indicate top (red) and bottom (blue) bilayer graphene, and interlayer WSe$_2$ (yellow). (c) $I_{int}$ vs $V_{int}$ at $V_{BG} = -20$ V for different $V_{TG}$ and $T$ values. The data shows resonance peaks and NDR that depend weakly on temperature.

Figure 1(b) shows an optical micrograph of a completed heterostructure. Multiple contacts to each graphene bilayer enable four-point tunneling current-voltage measurements, which decouple the intrinsic tunneling characteristics from the external resistances of the contacts and graphene access regions. The interlayer current ($I_{int}$) is measured as a function of the interlayer voltage ($V_{int}$) at fixed top gate ($V_{TG}$) and back gate ($V_{BG}$) voltages; $V_{int}$ is positive when the top layer is at a higher voltage with respect to bottom layer. Varying $V_{TG}$ and $V_{BG}$ tunes the total carrier density and its distribution between the layers. Changes in $V_{int}$ can also influence the distribution of charge between layers and directly alter the relative band structure alignment. Figure 1(c) shows a set of $I_{int}$ vs $V_{int}$ data at various $V_{TG}$ values and temperatures ($T$) at $V_{BG} = -20$ V. The data show clear tunneling resonance peaks and NDR that vary as a function of $V_{TG}$. We note that the $I_{int}$ vs $V_{int}$ data are largely insensitive to temperature, suggesting a minimal contribution from phonon assisted tunneling, and an interlayer tunneling energy that is insensitive to temperature.

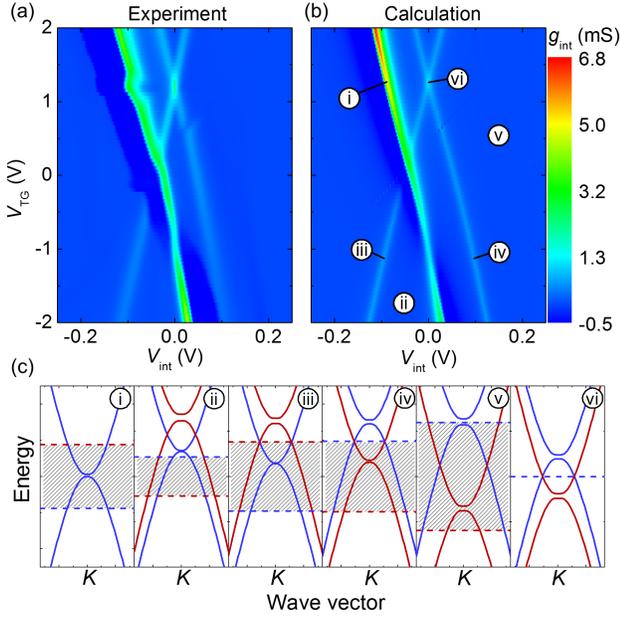

FIG. 2. Experimental (a) and calculated (b) $g_{int}$ vs $V_{int}$ and $V_{TG}$ at $V_{BG} = -20$ V and $T = 1.5$ K. Regions in which data are missing due to NDR circuit instabilities are linearly interpolated in panel (a). The points labeled in panel (b) identify distinct tunneling regimes. (c) Relative alignment of top (red) and bottom (blue) bilayer graphene bands corresponding to the biasing conditions labeled in panel (b). The dashed lines mark the two layers Fermi levels. At point (vi) the carrier densities are equal and opposite in the two layers.

To construct a picture of tunneling in double bilayer graphene-WSe$_2$ heterostructures, we consider the differential tunneling conductance ($g_{int} = \frac{dI_{int}}{dV_{int}}$) dependence on $V_{int}$ and $V_{TG}$, at $V_{BG} = -20$ V and $T = 1.5$ K, shown in Fig 2(a). The data show coupled lines of maximum $g_{int}$ and negative $g_{int}$, corresponding to the resonance and NDR conditions. There are two additional lines of increased $g_{int}$ forming an X pattern, similar to that of the resonance, that are discussed in more detail below.

To quantitatively understand Fig. 2(a) data, we employ a single-particle tunneling model to calculate $g_{int}$. The electrostatic potentials of both graphene bilayers are calculated self-consistently, including screening, to determine the relative band alignments and bandgap openings. The interlayer current is given by:

$$I_{int} = -e \int_{-\infty}^{\infty} T(E)(f(E - \mu_T) - f(E - \mu_B))dE \quad (1)$$

where $e$ is the electron charge, $E$ the energy, $f(E)$ the Fermi-Dirac distribution, and $\mu_T$ ($\mu_B$) is the top (bottom) layer Fermi level. The tunneling rate $T(E)$ is:

$$T(E) = \frac{2\pi}{\hbar} \sum_{k;ss'} |t|^2 A_{T,s}(k, E) A_{B,s'}(k, E) \quad (2)$$

The summation is over all momentum states ($k$) and the first two sub-bands ($s$ and $s'$) of the bilayer graphene conduction and valence bands. $A_{T,s}$ and $A_{B,s}$ are the spectral density functions of the band $s$ in the top and bottom bilayers, respectively, and $t$ is the interlayer coupling energy. The spectral density functions are Lorentzian in form:

$$A_s(k, E) = \frac{1}{\pi} \frac{\Gamma}{(E - \varepsilon_s(k))^2 + \Gamma^2} \quad (3)$$

$\varepsilon_s(k)$ is the bilayer graphene dispersion of band $s$, and $\Gamma$ is the quasiparticle state energy broadening. The $\varepsilon_s(k)$ dependence is computed using a simplified tight-binding model around the $K$-point, including the band gap opening in bilayer graphene in the presence of a transverse electric field [26, 27]. The only free parameters in the model are $t$ and $\Gamma$, which depend on disorder and the quality of the interfaces in the heterostructure. The values $t = 30$ $\mu$eV and $\Gamma = 4$ meV provide the best fit to the Fig. 2(a) data.

Figure 2(b) shows the calculated $g_{int}$ for the same $V_{int}$, $V_{TG}$, and $V_{BG}$ values in Fig. 2(a). A comparison of Fig. 2(a) and 2(b) reveals good agreement between the two datasets. Distinct tunneling regimes [labeled (i) - (vi) in Fig. 2(b)] are evident in both panels. The calculated band alignments at points (i) - (vi) are illustrated in Fig. 2(c). Line (i) corresponds to resonance, where the bands of the two bilayers fully align and a large density of states supports energy and momentum conserving tunneling. In region (ii), the bands are misaligned and energy and momentum conserving tunneling no longer occurs. Along lines (iii) and (iv), a ring of intersection occurring between an electron band of one bilayer and a hole band of the other bilayer crosses into the energy interval between the Fermi levels allowing energy and momentum conserving tunneling, referred to below as unlike-band tunneling. The symmetry of the unlike-band tunneling lines (iii) and (iv) is readily explained, since the ring of overlap will cross the Fermi level of opposite layers at opposite $V_{int}$ values at a given $V_{TG}$. In region (v), the ring of overlap lies between the two layer Fermi levels, allowing for unlike-band tunneling.

At point (vi), where the lines of unlike-band tunneling converge, the ring of overlap and both layer Fermi levels coincide, and the carrier density in the top bilayer graphene ($n_T$) is equal and opposite to the carrier density in the bottom bilayer graphene ($n_B$). Under this biasing condition, $V_{TG}$ and $V_{BG}$ balance each other, and the heterostructure is at total charge neutrality with a finite carrier density in each layer. This configuration is the most conducive to indirect exciton formation because an electron in one layer has a corresponding hole in the opposite layer at each momentum.

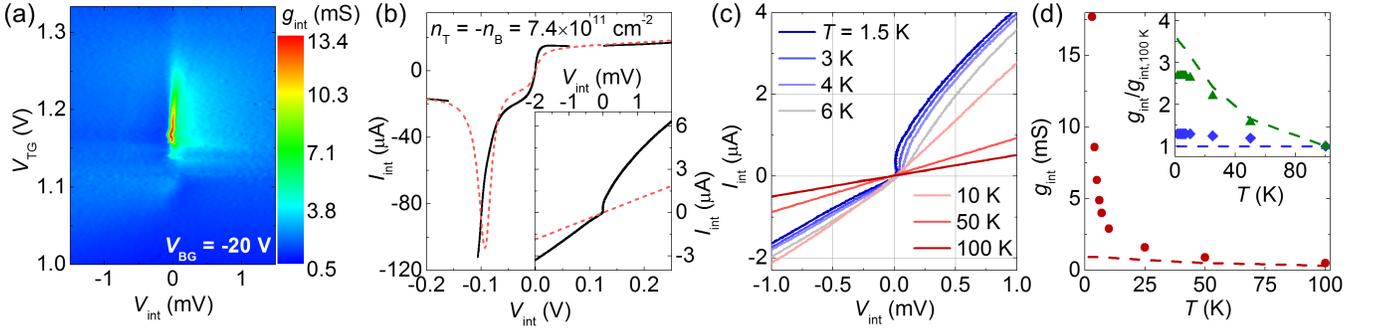

FIG. 3. (a) $g_{int}$ vs $V_{int}$ and $V_{TG}$ at $V_{BG}$ = -20 V and $T$ = 1.5 K, showing a strongly enhanced conductance in the vicinity of $n_T = -n_B$. The $g_{int}$ range is reduced to better compare regions of low and high $g_{int}$ values. (b) Experimental (solid black) and calculated (dashed red) $I_{int}$ vs $V_{int}$ at $n_T = -n_B$ ($V_{TG}$ = 1.18 V, $V_{BG}$ = -20 V), and $T$ = 1.5 K. The inset shows a magnified view of $I_{int}$ vs $V_{int}$ data near $V_{int}$ = 0 V. (c) $I_{int}$ vs $V_{int}$ at $n_T = -n_B = 7.4 \times 10^{11}$ cm$^{-2}$ measured at different temperatures, in the vicinity of $V_{int}$ = 0 V. (d) Measured (symbols) and calculated (dashed line) $g_{int}$ vs $T$ data at $V_{int}$ = 0 V and $n_T = -n_B = 7.4 \times 10^{11}$ cm$^{-2}$, showing a strong experimental $g_{int}$ increase with reducing $T$. Inset: temperature dependence of $g_{int}$ normalized to the $T$ = 100 K value, at resonance (blue diamonds) and at the onset of unlike-band tunneling (green triangles), for $n_T \approx n_B = -1.7 \times 10^{12}$ cm$^{-2}$, showing a weak temperature dependence in agreement with calculations (dashed lines).

We observe that the experimental $I_{int}$ and $g_{int}$ data depart significantly from the single-particle tunneling model at $n_T = -n_B$. Figure 3(a) shows a magnified view of the Fig. 2(a) data in the vicinity of $n_T = -n_B$, where a large peak in $g_{int}$ is visible at $V_{int}$ = 0 V. By comparison, the single-particle model does not distinguish $n_T = -n_B$ from any other point along a line of unlike-band tunneling onset, and does not predict an increase in $g_{int}$ at that point. Examining the experimental data, we see that the $g_{int}$ peak is very narrow with respect to $V_{int}$, suggesting a critical $I_{int}$ value beyond which the enhancement is reduced. Furthermore, at $n_T = -n_B$ the onset of $I_{int}$ vs $V_{int}$ is vertical within detection limits.

In Fig. 3(b) we show a comparison of experimental and calculated $I_{int}$ vs $V_{int}$ at $n_T = -n_B = 7.4 \times 10^{11}$ cm$^{-2}$. While the experimental data and calculations are in good agreement for most of the $V_{int}$ range, in the vicinity of $V_{int}$ = 0 V the experimental data exhibit a much sharper increase in $I_{int}$ compared to the single-particle model. The Fig. 3(b) inset shows a zoomed view of the same data near $V_{int}$ = 0 V, in which the experimental curve displays a vertical onset at $V_{int}$ = 0 V, and a differential conductance that is strongly enhanced compared to calculations.

Figure 3(c) shows the $I_{int}$ vs $V_{int}$ data at temperatures between $T$ = 1.5 K and 100 K, for the same biasing conditions as in Fig. 3(b). The enhanced tunneling weakens rapidly with increasing $T$, and the vertical $I_{int}$ vs $V_{int}$ onset is suppressed by $T$ = 10 K. Above $T$ = 50 K the $I_{int}$ vs $V_{int}$ dependence becomes linear. We note that the Fig. 3(b-c) data are not symmetric with respect to $V_{int}$ = 0 V, and small layer density imbalances can shift the enhanced $I_{int}$ onset from positive to negative values (Fig. S1). The origin of this asymmetry is unclear at present.

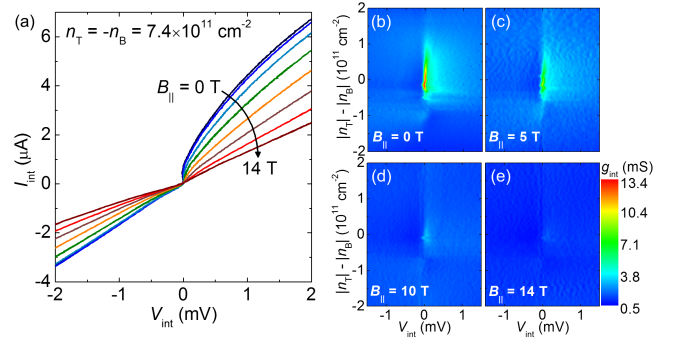

FIG. 4. (a) $I_{int}$ vs $V_{int}$ measured at $n_T = -n_B = 7.4 \times 10^{11}$ cm$^{-2}$, and at different $B_{||}$ values, from 0 T to 14 T in steps of 2 T. (b-e) $g_{int}$ vs $V_{int}$ and $|n_T| - |n_B|$ near $n_T = -n_B$, at different $B_{||}$ values. The $g_{int}$ enhancement is significantly suppressed in the presence of an in-plane magnetic field.

Figure 3(d) shows the maximum experimental (symbols) and calculated (dashed line) $g_{int}$ values as a function of $T$. The experimental $g_{int}$ rises sharply with decreasing $T$, in contrast with the weak temperature dependence in the single-electron theoretical model. The measured data point at $T$ = 1.5 K is not included because the vertical $I_{int}$ vs $V_{int}$ onset renders $g_{int}$ very large, a regime where current spreading within the heterostructure can impact the measurement accuracy. Figure 3(d) inset compares the $T$ dependence of experimental (symbols) and calculated (dashed lines) $g_{int}$ curves normalized to their $T$ = 100 K values at the onset of unlike-band tunneling (green) and at resonance (blue) for layer densities $n_T \approx n_B = -1.7 \times 10^{12}$ cm$^{-2}$ at $V_{int}$ = 0 V. Unlike the behavior at total charge neutrality, both sets of data show relatively weak temperature dependence, similar to Fig. 1(c), and match closely with the dependence predicted by the single-particle calculations, where the Fermi-Dirac distribution broadening controls the temperature dependence.

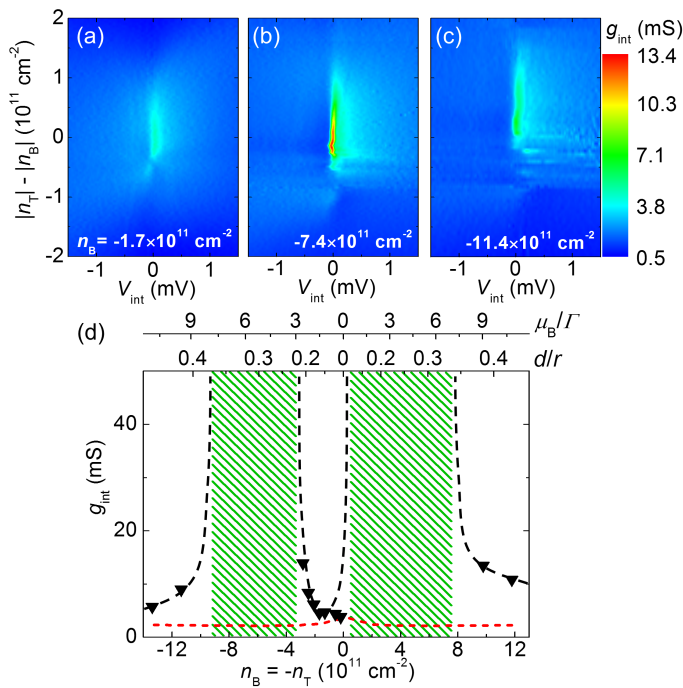

FIG. 5. (a-c) $g_{int}$ vs $V_{int}$ and $|n_T| - |n_B|$ near $n_T = -n_B$, at different $n_B$. (d) Experimental (black triangles) and calculated (red dashed line) $g_{int}$ vs $n_B$, at $n_T = -n_B$. The shaded regions indicate densities at which the experimental $I_{int}$ onset is vertical to within experimental accuracy at $V_{int} = 0$ V. The black dashed lines are guides to the eye. The lower (upper) top x-axis shows the $d/r$ ($\mu_B/\Gamma$) ratio.

While broadening at the Fermi level plays a role in the temperature dependence at total neutrality, it is not sufficient to explain the sharp increase in $g_{int}$ below $T = 25$ K. The enhanced tunneling at $n_T = -n_B$ instead suggests the emergence of a many-body state associated with the formation of indirect excitons across the two graphene bilayers. In analogy with previous experimental [5] and theoretical [28, 29] studies of quantum Hall exciton superfluids, we interpret the experimental observations as evidence for the presence of electron-hole pairs that effectively short the two graphene bilayers, allowing carriers to recombine without dissipation. The spatial coherence of the interlayer phase can be measured directly by applying a magnetic field ($B_{||}$) along the x-direction of the sample $x - y$ plane to add a phase factor $e^{i(2\pi y/L)}$ to the interlayer tunneling amplitudes, where $L = h/edB_{||}$ and $d = 2.2$ nm is the interlayer distance in our heterostructure. As illustrated in Fig. 4(a) the $I_{int}$ vs $V_{int}$ data approach linearity and the enhanced tunneling is suppressed at $B_{||} = 10$ T [Fig. 4(b-e)], implying a phase coherence length of $\approx 0.2$ $\mu$m [29–31].

Next, we discuss the layer density dependence of the tunneling characteristics at $n_T = -n_B$. Suppression of exciton condensation due to intralayer screening of interlayer Coulomb interactions has been predicted at large densities, as well as suppression due to disorder or competing phases at small densities [14, 32]. In Fig. 5(a-c), we show $g_{int}$ as a function of $V_{int}$ and $|n_T| - |n_B|$ at $n_B = -1.7 \times 10^{11}$ cm$^{-2}$, $-7.4 \times 10^{11}$ cm$^{-2}$, and $-11.4 \times 10^{11}$ cm$^{-2}$, respectively. All three datasets exhibit enhanced $g_{int}$ at $n_T = -n_B$, but the enhancement is greatly reduced for the smallest and largest $n_B$, with both reaching a maximum $g_{int}$ of $\sim$5 mS. The intermediate $n_B$ data are the same as in Fig. 3(a), which show a divergent $g_{int}$ at total charge neutrality. Examples of $g_{int}$ vs $V_{int}$ data at total charge neutrality at different $n_B$ and $T$ values are included in the supplementary material (Fig. S2).

Figure 5(d) shows the maximum experimental and calculated $g_{int}$ vs $n_B$ near $n_T = -n_B$. The shaded regions indicate densities at which the measured onset of $I_{int}$ at $V_{int} = 0$ V is vertical and $g_{int}$ nominally diverges. To characterize the density dependence we examine the ratio of the interlayer distance to the average interparticle spacing $r = 1/\sqrt{\pi|n_B|}$. The $d/r$ values are shown on the lower top x-axis of Fig. 5(d). The data reveal that the tunneling enhancement is reduced for $d/r \geq 0.35$, consistent with theoretical considerations of intralayer correlations overcoming the interlayer pairing when $r$ is comparable to $d$ [32, 33]. Figure 5(d) upper top x-axis shows the $\mu_B/\Gamma$ ratio, which suggests that at small densities, disorder precludes exciton formation when the quasiparticle state energy broadening is comparable to, or larger than the layer Fermi level. The asymmetry of the $g_{int}$ enhancement with $n_B$ may be related to differences in disorder between the two layers.

Lastly we comment on the sample design, and in particular the use of WSe$_2$ as tunnel barrier. Because the tunneling conductance enhancement due to interlayer coherence is expected to scale as $t^2$ [30, 31] above the ordering temperature, the effect is not easily probed when the tunneling barrier is either extremely transparent or extremely opaque. The WSe$_2$ tunnel barrier satisfies this requirement, has a favorable band alignment with graphene [24], and high crystal quality [22].

The observation of strongly enhanced tunneling between bilayer graphene samples separated by WSe$_2$ points towards the presence of an emerging many-body state with electron-hole pair condensation. A single-particle tunneling model accurately predicts tunneling characteristics except at overall neutrality. Further theoretical work is needed to fully explain the tunneling behavior at total charge neutrality, and the nonlinear $I_{int}$ vs $V_{int}$ dependence at low temperatures.

We thank Dmitri Efimkin and Javad Shabani for discussions. This work was supported by National Science Foundation grant EECS-1610008, the Nanoelectronics Research Initiative SWAN center, and the Army Research Office under Award W911NF-17-1-0312. K.W. and T.T. acknowledge support from the Elemental Strategy Initiative conducted by the MEXT, Japan and JSPS KAKENHI Grant Numbers JP15K21722.

---

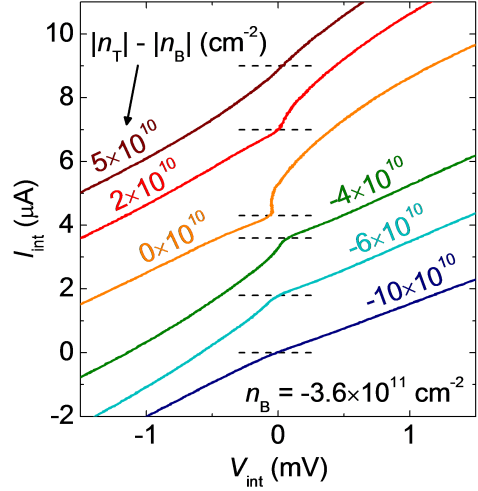

FIG. S1. $I_{int}$ vs $V_{int}$ at various density imbalances between the two graphene bilayers ($|n_T| - |n_B|$) for an initial $n_B = -3.6\times10^{11}$ cm$^{-2}$. The curves are vertically offset for clarity. The horizontal dashed lines mark the $I_{int} = 0$ value for each trace. A vertical tunneling characteristic is observed at $V_{int} = 0$ V near zero imbalance and decreases with an increasing $|n_T| - |n_B|$ magnitude. For negative density imbalance, the tunneling enhancement changes polarity, i.e. the step feature in $g_{int}$ occurs at negative $I_{int}$ values.

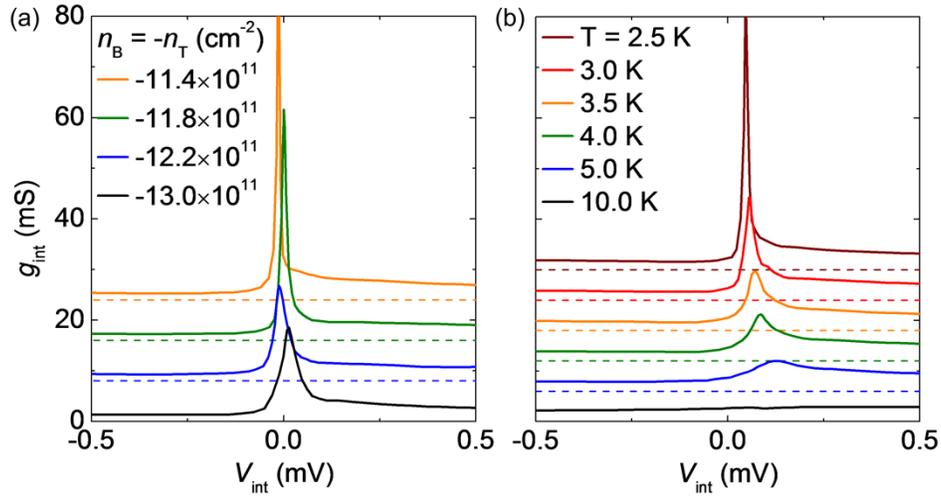

FIG. S2. (a) $g_{int}$ vs $V_{int}$ at $n_T = -n_B$ for different $n_B$ values, showing a suppression of the enhanced tunneling with increasing density. The behavior at low densities, when $|n_B|$ approaches zero is similar. (b) $g_{int}$ vs $V_{int}$ at $n_T = -n_B$ for $n_B = -5.5\times10^{11}$ cm$^{-2}$ at different temperatures. The conductance peak rapidly decreases with increasing temperature. The traces in panels (a) and (b) are offset for clarity. The dashed lines indicate $g_{int} = 0$ for each trace. We note that panel (a-b) data were measured in a cooldown different than Figs. 3 and 5 data, and the conductance values differ slightly from the main text figures.